\documentclass[conference]{IEEEtran}
\IEEEoverridecommandlockouts
\usepackage{amsmath,amssymb,amsfonts}
\usepackage{algorithmic}
\usepackage[ruled]{algorithm2e}
\usepackage{array}
\usepackage{textcomp}
\usepackage{stfloats}
\usepackage{url}
\usepackage{color}
\usepackage{verbatim}
\usepackage{graphicx}
\usepackage{tabularx}
\usepackage{multirow}
\usepackage{array}
\usepackage{booktabs}
\usepackage{graphicx}
\usepackage{parskip}

\newtheorem{definition}{Definition}
\usepackage[numbers]{natbib}
\usepackage{bm}
\usepackage{mathtools}

\usepackage{subfigure}

\usepackage{bibspacing}
\graphicspath{ {./fig/} }
\usepackage{enumitem}

\def\BibTeX{{\rm B\kern-.05em{\sc i\kern-.025em b}\kern-.08em
    T\kern-.1667em\lower.7ex\hbox{E}\kern-.125emX}}
\begin{document}

\title{Accurate Trust Evaluation for Effective Operation of Social IoT Systems via Hypergraph-Enabled Self-Supervised Contrastive Learning\\
}

\author{\IEEEauthorblockN{Botao Zhu and Xianbin Wang}
\IEEEauthorblockA{Dept. of Electrical and Computer Engineering, Western University,
London, Ontario N6A 3K7 CANADA \\}   

}

\maketitle

\begin{abstract}

Social Internet-of-Things (IoT) enhances collaboration between devices by endowing IoT systems with social attributes. However, calculating trust between devices based on complex and dynamic social attributes—similar to trust formation mechanisms in human society—poses a significant challenge. To address this issue, this paper presents a new hypergraph-enabled self-supervised contrastive learning (HSCL) method to accurately determine trust values between devices. To implement the proposed HSCL, hypergraphs are first used to discover and represent high-order relationships based on social attributes. Hypergraph augmentation is then applied to enhance the semantics of the generated social hypergraph, followed by the use of a parameter-sharing hypergraph neural network to nonlinearly fuse the high-order social relationships. Additionally, a self-supervised contrastive learning method is utilized to obtain meaningful device embeddings by conducting comparisons among devices, hyperedges, and device-to-hyperedge relationships. Finally, trust values between devices are calculated based on device embeddings that encapsulate high-order social relationships. Extensive experiments reveal that the proposed HSCL method outperforms baseline algorithms in effectively distinguishing between trusted and untrusted nodes and identifying the most trusted node.

\end{abstract}


\section{Introduction}

Social Internet-of-Things (IoT) is an emerging paradigm which integrates social networking and interaction principles into IoT systems to enhance the cooperation among devices~\cite{b1}. By empowering IoT devices with the new capability of autonomous social relationship management, much like humans interact with others in social networks, social IoT can dramatically enhance functionality and over-system performance under diverse operational conditions.

Fulfilling the overall objective of social IoT relies on trusted information sharing, collaboration, and decision-making in complex IoT systems with changing operational conditions. One key challenge here is the issue of dynamic trust evaluation with changing operational objectives and environments, which is critical for ensuring effective cooperation among devices and ultimately enhancing the overall operation of social IoT systems.
Trust is generally viewed as the belief or confidence a trustor has in a trustee's ability to complete a task that meets the trustor's expectations within a given situation~\cite{b2}. In social IoT, the social attributes among devices, such as spatial location, historical interactions, reputation, and shared interests, are considered essential factors influencing trust. Various approaches have been proposed to infer trust among devices by analyzing these social attributes, which can be broadly classified into three categories: 


\subsubsection{Linear weighted sum methods} 

These techniques first calculate the similarity between devices based on each social attribute and then compute the weighted sum of all these similarities to determine the trust values between devices~\cite{b3}.


\subsubsection{Matrix methods} 
A series of device-to-device attribute matrices are constructed based on factors like historical behaviour, preferences, and characteristics~\cite{b5} and then apply matrix operations, such as matrix factorization~\cite{b6}, to predict trust between devices.



\subsubsection{Machine learning methods} 
The related methods generally start by collecting social information about devices, such as behaviour and user interactions, followed by labelling the data through various methods, and finally training machine learning models on the labeled data to predict trust values~\cite{b7}.

However, the aforementioned trust evaluation methods fail to emulate the intricate trust mechanisms in human society, such as the interactions among complex relationships and their nonlinear integration. As a result, they cannot accurately reflect the true trust between devices in IoT systems characterized by dynamic and complex social relationships. Specifically, linear weighted sum methods treat the similarity of each social attribute between devices as one aspect of trust and calculate overall trust simply as the linear sum of a series of point-to-point relationships. Similarly, matrix-based methods also account for a series of point-to-point attribute relationships, but matrix operations can become intractable as the system scales with more devices and social attributes. Additionally, machine learning approaches require large amounts of labeled data to train trust inference models; however, the diversity and dynamic nature of social attribute relationships between devices make it challenging to obtain sufficient labeled data.

Given the limitations of existing approaches, a new method that emulates trust formation mechanisms in human society is urgently required to accurately assess the true trust between devices in social IoT systems. This method should overcome the constraints of simple point-to-point attribute relationships and address the reliance on labeled attribute data. Due to their strength in representing complex relationships~\cite{bzhu}, hypergraphs are particularly suitable for modelling social relationships between devices. Therefore, we propose a new trust evaluation technique based on the hypergraph-enabled self-supervised contrastive learning (HSCL), which captures high-order social relationships that go beyond point-to-point interactions, performs a nonlinear fusion of these relationships, and utilizes an unsupervised training method. This technique allows any device to easily identify the most trustworthy collaborator based on social attributes. The main contributions of this paper are summarized below. 

\begin{itemize}[leftmargin=*]
    \item We creatively employ hypergraphs to extract and represent high-order relationships based on social attributes in social IoT systems, accurately capturing mutual influences among a group of devices beyond traditional pairwise interactions.

    \item To model the combined effects of a set of high-order social relationships on trust between devices, hypergraph augmentation is utilized to enrich the semantics of the generated social relationship hypergraph, while a parameter-sharing hypergraph neural network (HGNN) is employed to nonlinearly fuse the high-order social relationships within the augmented hypergraphs. 

    \item To effectively calculate trust between devices, a self-supervised contrastive learning approach is employed to learn device embeddings that integrate high-order social relationships from the social relationship hypergraph. Subsequently, the trust values between devices are calculated based on the obtained embeddings.

    \item Extensive experiments demonstrate that the proposed HSCL method can clearly distinguish between trusted and untrusted nodes and select the most trusted device compared to the baseline algorithms.
    
\end{itemize}


\section{System Model and Problem Definition}
\label{problem}
This paper considers a social IoT system consisting of a set of devices, defined as $\bm{A} = \{a_1,\dots,a_I\}$. These devices are interconnected with a certain set of social attributes, and all social attributes are defined as a set $\bm{S}$. Devices assess their mutual trust levels based on their social attributes and establish trustworthy cooperative relationships. Therefore, accurately defining trust in the social IoT system is a crucial prerequisite for trust evaluation between devices. We define trust as follows:
\begin{definition}[Trust in the social IoT system]
   \textit{For any pair of devices $a_i, a_j \in \bm{A}$ in the social IoT system, the trust of device $a_i$ in $a_j$ is the likelihood that $a_j$ can assist $a_i$ considering the entire system and the social attributes between them, which is given by}
   \vspace{-0.05 in}
   \begin{align}
       T_{a_i \to a_j} = TRUST(a_i,a_j,\bm{A}, \bm{S}). 
   \end{align}
\end{definition}
We can see that trust between any pair of devices is determined by the collective influence of all devices in the system and the social attributes connecting them. Similar to evaluating a person's trustworthiness in human social systems, we always comprehensively consider the evaluations of other people from different communities. The purpose of calculating trust between devices is to assist them in identifying reliable collaborators. If $a_i$ is the task initiator seeking the most trustworthy device in the system for collaboration, the problem of identifying the most trusted collaborator can be expressed as follows:
\begin{align}
    \arg \max_{a_j \in \bm{A}, a_j \neq a_i} TRUST(a_i, a_j, \bm{A}, \bm{S}).
\end{align}
The key to solving this problem is accurately assessing the trust values of all potential collaborators. Once their trust values are obtained, $a_i$ can easily select the collaborator with the highest trust value. To achieve this goal, we propose a trust calculation model based on the HSCL method.



\section{Trust Evaluation Model Based on Hypergraph-Enabled Self-Supervised Contrastive Learning}
\label{hscl}





\begin{figure}[!]
\centering
\includegraphics[scale=0.7]{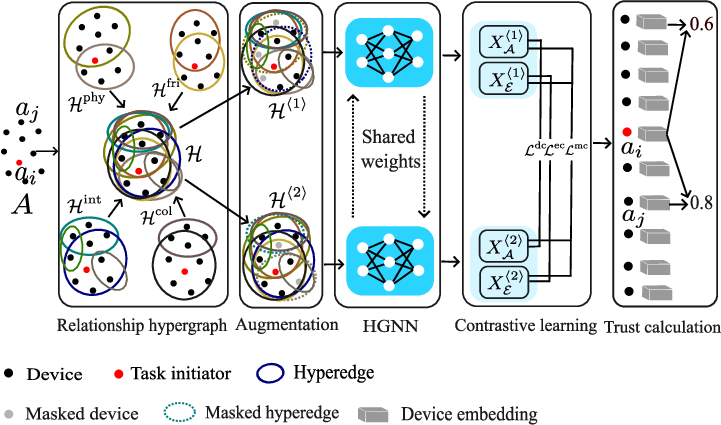}
\caption{The framework of the proposed HSCL.}
\label{framework}
\end{figure}

Existing trust computation methods in social IoT systems frequently overlook critical aspects, such as high-order social relationships among multiple devices and the seamless integration of these relationships. As a result, the computed trust values often fail to accurately represent the genuine trust between devices. In this study, the proposed HSCL method overcomes these limitations. First, hypergraphs are employed to mine and represent the complex and high-order relationships based on social attributes within the system. In addition, hypergraph augmentation, HGNN, and self-supervised contrastive learning are utilized to learn the embeddings of devices that incorporate complex social relationships. Finally, trust values between devices are calculated based on the obtained embeddings. The framework of the proposed HSCL method is shown in Fig.~\ref{framework},  with the details outlined below.


\subsection{Hypergraph-driven high-order social relationship representation}
In this subsection, the basic concepts of hypergraphs are first introduced. Then, the physical spatial attribute $s^{\text{phy}}$, the friendship attribute $s^{\text{fri}}$, the community-of-interest (CoI) attribute $s^{\text{int}}$, and the collaboration attribute $s^{\text{col}}$ are considered as social attributes $\bm{S} = \{s^{\text{phy}}, s^{\text{fri}}, s^{\text{int}}, s^{\text{col}}\}$,  and hypergraphs are utilized to establish high-order, nonlinear social relationships between devices based on these attributes.

\subsubsection{Hypergraph}
A hypergraph $\mathcal{H}$ is denoted as $\mathcal{H} = (\mathcal{A}, \mathcal{E})$, where $\mathcal{A}$ is the set of all nodes, and $\mathcal{E}$ is the set of all hyperedges. Each hyperedge $e$ can contain a certain number of nodes, representing the relationship between these nodes. Each node $a$ can form different relationships with other nodes. The hypergraph structure can be represented by an incidence matrix $\bm{H} \in \mathbb{R}^{|\mathcal{A}| \times |\mathcal{E}|}$, with entries $h(a, e)$ defined as
\vspace{-0.07 in}
 \begin{align}
    h(a,e) = \begin{cases}
    1, & a \in e;\\
    0, & a \notin e.
   \end{cases}
\end{align}

\subsubsection{Physical spatial relationship}
$s^{\text{phy}}$ represents the proximity between devices in the physical space. To accurately capture the spatial relationships between devices, the soft $K$-means clustering algorithm is employed, which can group devices that are close in physical space into the same cluster and allow a device to belong to multiple clusters~\cite{b9}. We first randomly select $K$ devices from $\bm{A}$ as the initial center nodes $\{a_1^{\text{cn}}, \dots, a_K^{\text{cn}}\}$ of the soft $K$-means, and then minimize the cost function
\vspace{-0.1 in}
\begin{align}
    \min \sum_{k=1}^{K}\sum_{i=1}^{I} z_{ki} ||a_i - a_k^{\text{cn}}||^2,
\end{align}
where $||a_i - a^{\text{cn}}_k||$ is Euclidean distance between $a_i$ and $a^{\text{cn}}_k$, $z_{ki}$ is the membership probability of device $a_i$ to the $k$-th center node. $z_{ki}$ is defined as 
\vspace{-0.1 in}
\begin{align}
    z_{ki} = \frac{e^{-\beta {||a_i - a_k^{\text{cn}}||}^2}}{\sum_{k=1}^{K}e^{-\beta {||a_i - a_k^{\text{cn}}||}^2}}, 
\end{align}
where $\beta$ is the stiffness parameter, and $z_{ki} \in [0, 1]$, $\sum_{k=1}^{K}z_{ki} = 1$. The soft $K$-means iteratively updates $z_{ki}$ of each device and the cluster center nodes. Each cluster center node in each iteration is calculated by
\begin{align}
    a^{\text{cn}}_{k} = \frac{\sum_{i=1}^{I}z_{ki}a_i}{\sum_{i=1}^{K}z_{ki}}. 
\end{align}
Until the soft $K$-means algorithm converges, a set of clusters $\{C_1,\dots, C_K \}$ is obtained. Each cluster $C_k$ is then enclosed by an edge $e^{\text{phy}}_k$, which serves as a hyperedge representing the physical spatial relationships among devices within $C_k$. Finally, the physical spatial hypergraph $\mathcal{H}^{\text{phy}} = \{e^{\text{phy}}_{1}, \dots, e^{\text{phy}}_{K} \}$ is obtained by combining all $K$ hyperedges.  

\subsubsection{CoI relationship} $s^{\text{int}}$ represents a group of devices having a common interest in the same subject. Devices with the same $s^{\text{int}}$ are more likely to collaborate with each other. Assuming the total number of interests in the system is $B$, each interest $b \in B$ can be regarded as a hyperedge encapsulating all devices sharing $b$, denoted as $e^{\text{int}}_b$. Ultimately, all hyperedges collectively form the interest hypergraph $\mathcal{H}^{\text{int}} = \{e^{\text{int}}_{1},  \dots, e^{\text{int}}_{B}$\}.

\subsubsection{Friendship} 
$s^{\text{fri}}$ plays an important role in inferring trust between devices. If a group of devices shares a friendship, they exhibit a high level of mutual trust. This friendship is represented by a hyperedge, denoted as $e^{\text{fri}}$, which encompasses all devices in the group. Assuming there are $G$ friendships, the hyperedges formed by these relationships collectively create a friendship hypergraph $\mathcal{H}^{\text{fri}} = \{e^{\text{fri}}_1,\dots, e^{\text{fri}}_{G}\}$.

\subsubsection{Collaborative relationship} $s^{\text{col}}$ reflects the past collaborations among devices, indicating the potential for future cooperation. If a group of devices have collaborated, a hyperedge is used to encapsulate them, representing their collaborative relationship. The weight of each hyperedge represents the effectiveness of the collaboration, with 1 indicating successful collaboration and 0 indicating failure. Assuming there are $F$ collaborative relationships, the hyperedges generated by these relationships collectively form a collaborative hypergraph $\mathcal{H}^{\text{col}} = \{e^{\text{col}}_1,\dots, e^{\text{col}}_F\}$.

To integrate all social relationships, $\mathcal{H}^{\text{phy}}$, $\mathcal{H}^{\text{int}}$, $\mathcal{H}^{\text{fri}}$, and $\mathcal{H}^{\text{col}}$ are concatenated to form a social relationship hypergraph $\mathcal{H} = \left(\mathcal{A}, \mathcal{E} \right)$, where $\mathcal{A} = \bm{A}$, and $\mathcal{E}$ is the set of all hyperedges. To unify the notation, all hyperedges are re-expressed as $\mathcal{E} = \{e_n \}_{n=1}^{|\mathcal{E}|}$. Each hyperedge is associated with a weight $w_n$, and the matrix of weights is $\bm{W} \in \mathbb{R}^{|\mathcal{E}| \times |\mathcal{E}|}$. The feature matrix of all devices in $\mathcal{H}$ is represented as $\bm{X}_{\mathcal{A}} \in \mathbb{R}^{|\mathcal{A}| \times d}$, and the incidence matrix of $\mathcal{H}$ is $\bm{H} \in \mathbb{R}^{|\mathcal{A}| \times |\mathcal{E}|}$. The degree of devices is denoted by the diagonal matrix $\bm{D}_{a} \in \mathbb{R}^{|\mathcal{A}| \times |\mathcal{A}|}$, where each element $\delta(a_i) = \sum_{e_n \in \mathcal{E}} w_{n} h(a_i,e_n)$. The degree of hyperedges is denoted by the diagonal matrix $\bm{D}_{e} \in \mathbb{R}^{|\mathcal{E}| \times |\mathcal{E}|}$, where each element $\delta(e_n) = \sum_{a_i \in e_n}h(a_i,e_n)$ representing the number of devices connected by $e_n$.

\subsection{Hypergraph-enabled self-supervised contrastive learning}

$\mathcal{H}$ captures various social relationships between devices, but trust between any pair of devices cannot be directly calculated from these relationships. To solve this issue, it is necessary to learn the devices and their social relationships and map them into a space of the same dimension. Therefore, hypergraph learning is used to learn a mapping function $f_\theta: \mathcal{H} \to (\bm{X}_{\mathcal{A}}, \bm{X}_{\mathcal{E}})$, where $\bm{X}_{\mathcal{A}}$ and $\bm{X}_{\mathcal{E}}$ are the embeddings of devices and hyperedges, respectively. To train $f_\theta$, the self-supervised contrastive learning method is utilized, which has excelled in computer vision by learning data representations directly from raw data~\cite{b11}. This technique begins by creating two augmented views from raw data to provide different contexts or semantics, then learns a machine learning model to maximize the agreement between these views. The learning architecture primarily consists of three components: hypergraph augmentation, hypergraph embedding, and optimization for self-supervised contrastive objectives.

\subsubsection{Hypergraph augmentation} 
To create two augmented views, three types of data augmentation are utilized: device masking, hyperedge masking, and device-hyperedge membership masking. For device masking, a vector $\bm{M}^{\mathcal{A}} \in \{0, 1\}^{|\mathcal{A}|}$ is constructed, where each element is independently drawn from a Bernoulli distribution $\mathcal{B}(1 - p^{\mathcal{A}})$, with $p^{\mathcal{A}}$ being the drop probability of devices. For hyperedge masking, we sample a vector $\bm{M}^{\mathcal{E}} \in \{0, 1\}^{|\mathcal{E}|}$ from a Bernoulli distribution $\mathcal{B}(1 - p^{\mathcal{E}})$, where $p^{\mathcal{E}}$ is the drop probability of hyperedges. For device-hyperedge masking, a masking matrix {\footnotesize $\bm{M}^{\bm{H}} \in \{0, 1\}^{|\mathcal{A}| \times |\mathcal{E}|}$} is constructed, where each entry is sampled from a Bernoulli distribution $\mathcal{B}(1 - p^{\bm{H}})$, with $p^{\bm{H}}$ being the drop probability of device-hyperedge membership links. The augmented view 1, $\mathcal{H}^{\langle 1 \rangle}$, is computed as follows:
\begin{align}
    \mathcal{A}^{\langle 1 \rangle} =  \mathcal{A} \odot \bm{M}^{\mathcal{A}},
    \mathcal{E}^{\langle 1 \rangle} = \mathcal{E} \odot \bm{M}^{\mathcal{E}},
    \bm{H}^{\langle 1 \rangle} = \bm{H} \odot \bm{M}^{\bm{H}},
\end{align}
where $\odot$ is the element-wise multiplication. Similarly, the augmented view 2, $\mathcal{H}^{\langle 2 \rangle}$, can be obtained using the same equations. The degree of augmentation is controlled by $p^{\mathcal{A}}$, $p^{\mathcal{E}}$, and $p^{\bm{H}}$.

\vspace{0.06 in}
\subsubsection{Hypergraph embedding} To produce device and hyperedge embeddings for the two augmented views, a parameter-sharing HGNN is utilized. Each view is inputted into HGNN, which employs a two-stage neighbourhood aggregation scheme: device-to-hyperedge and hyperedge-to-device. HGNN iteratively updates the representation of each hyperedge by aggregating representations of its incident devices, which is given by
\vspace{-0.1 in}
\begin{align}
\label{etoa}
    \bm{x}^{(l)}_{e_n} = f^{(l)}_{\mathcal{A} \to \mathcal{E}} \left(\bm{x}^{(l-1)}_{e_n}, \left \{ \bm{x}^{{(l-1)}}_{a_i} : a_i \in e_n \right \}\right),
\end{align}
where $\bm{x}^{(l-1)}_{e_n}$ and $\bm{x}^{{(l-1)}}_{a_i}$ are the embeddings of $e_n$ and $a_i$ at layer $(l - 1)$, respectively. In addition, the representation of each device is updated iteratively through aggregating representations of its incident hyperedges
\begin{align}
    \label{atoe}
    \bm{x}^{(l)}_{a_i} = f^{(l)}_{\mathcal{E} \to \mathcal{A}} \left(\bm{x}^{(l - 1)}_{a_i}, \left \{ \bm{x}^{(l)}_{e_n}: a_i \in e_n \right \} \right).
\end{align}
Formally, equations (\ref{etoa}) and (\ref{atoe}) in the $l$-th layer of HGNN can be represented in matrix form
\begin{align}
    \bm{X}^{(l)}_{\mathcal{E}} &=  \phi\left (\bm{D}^{-1}_{e} \bm{H}^T 
    \bm{X}^{(l - 1)}_{\mathcal{A}} \bm{\Theta}^{(l)}_{\mathcal{E}} \right ), \\
    \bm{X}^{(l)}_{\mathcal{A}} &= \phi \left (\bm{D}^{-1}_a \bm{H} \bm{W} \bm{X}^{(l)}_{\mathcal{E}} \bm{\Theta}^{(l)}_{\mathcal{A}} \right), 
\end{align}
where $\bm{X}^{(l)}_{\mathcal{E}} \in \mathbb{R}^{|\mathcal{E}| \times d}$ and $\bm{X}^{(l)}_{\mathcal{A}} \in \mathbb{R}^{|\mathcal{A}| \times d}$ are hyperedge and device embeddings at the $l$-th layer. $d$ is the embedding dimensionality. $ \bm{\Theta}^{(l)}_{\mathcal{E}}$ and $\bm{\Theta}^{(l)}_{\mathcal{A}}$ are trainable parameters for $f^{(l)}_{\mathcal{A} \to \mathcal{E}}$ and $f^{(l)}_{\mathcal{E} \to \mathcal{A}}$, respectively. $\phi(\cdot)$ denotes a nonlinear activation function. The initial input $\bm{X}^{(0)}_{\mathcal{A}}$ is $\bm{X}_{\mathcal{A}}$. Finally, HGNN outputs the embeddings of devices and hyperedges, $(\bm{X}^{{\langle 1 \rangle}}_{\mathcal{A}}, \bm{X}^{{\langle 1 \rangle}}_{\mathcal{E}})$ and $(\bm{X}^{{\langle 2 \rangle}}_{\mathcal{A}}, \bm{X}^{{\langle 2 \rangle}}_{\mathcal{E}})$, for $\mathcal{H}^{\langle 1 \rangle}$ and $\mathcal{H}^{\langle 2 \rangle}$, respectively.

\vspace{0.06 in}
\subsubsection{Optimization for self-supervised contrastive objectives} 
To learn more meaningful embeddings, three contrastive objectives are utilized: device contrast, hyperedge contrast, and device-hyperedge membership contrast.

\textit{Device contrast} aims to distinguish the representations of the same devices in the two augmented views from the representations of other devices. For any device $a_i \in \bm{A}$, its embedding from the first view, denoted as $\bm{x}^{\langle 1 \rangle}_{a_i} \in \bm{X}^{\langle 1 \rangle}_{\mathcal{A}}$, is set to the anchor, while its embedding from the second view, denoted as $\bm{x}^{\langle 2 \rangle}_{a_i} \in \bm{X}^{\langle 2 \rangle}_{\mathcal{A}}$, is treated as the positive sample. The embeddings of other devices from the second view are regarded as negative samples. Cosine function $\cos()$ is used to calculate the similarity of embeddings between two views, where the positive pair is assigned a high value and the negative pair is assigned a low value. The InfoNCE loss is utilized \cite{b12}, the loss function of each positive device pair between two views is given by
\vspace{-0.1 in}
\begin{align}
    \ell \left(\bm{x}^{\langle 1 \rangle}_{a_i}, \bm{x}^{\langle 2 \rangle}_{a_i} \right) = - \log \frac{e^{\cos \left(\bm{x}^{\langle 1 \rangle}_{a_i}, \bm{x}^{\langle 2 \rangle}_{a_i}\right)/ \tau^{\text{dc}}}}{\sum_{j=1}^{|\mathcal{A}|} e^{\cos \left(\bm{x}^{\langle 1 \rangle}_{a_i}, \bm{x}^{\langle 2 \rangle}_{a_j} \right)/ \tau^{\text{dc}}}},
\end{align}
where $\tau^{\text{dc}}$ is a temperature parameter. The objective function for device contrast is defined as the average loss across all positive pairs
\vspace{-0.1 in}
\begin{align}
    \mathcal{L}^{\text{dc}} = \frac{1}{2|\mathcal{A}|} \sum^{|\mathcal{A}|}_{i=1} \left( \ell \left(\bm{x}^{\langle 1 \rangle}_{a_i}, \bm{x}^{\langle 2 \rangle}_{a_i} \right) + \ell \left(\bm{x}^{\langle 2 \rangle}_{a_i}, \bm{x}^{\langle 1 \rangle}_{a_i} \right) \right).
\end{align}
\textit{Hyperedge contrast} focuses on differentiating the embedding of the same hyperedge across two augmented views from other hyperedge embeddings, aiding the model in retaining hyperedge information within the hypergraph. For any hyperedge $e_{n} \in \mathcal{E}$, its embedding from the first view, $\bm{x}^{\langle 1 \rangle}_{e_{n}}$, is set to the anchor, while its corresponding embedding in the second view, $\bm{x}^{\langle 2 \rangle}_{e_{n}}$, is treated as the positive sample, with all other embeddings from the second view considered as negative samples. Similar to device contrast, the loss function for each positive hyperedge pair is defined as 
\begin{align}
    \ell \left(\bm{x}^{\langle 1 \rangle}_{e_{n}}, \bm{x}^{\langle 2 \rangle}_{e_{n}} \right) = - \log \frac{e^{\cos \left(\bm{x}^{\langle 1 \rangle}_{e_{n}}, \bm{x}^{\langle 2 \rangle}_{e_{n}}\right)/ \tau^{\text{ec}}}}{\sum_{k=1}^{|\mathcal{E}|} e^{\cos \left(\bm{x}^{\langle 1 \rangle}_{e_{n}}, \bm{x}^{\langle 2 \rangle}_{e_{k}} \right)/ \tau^{\text{ec}}}},
\end{align}
where $\tau^{\text{ec}}$ is a temperature parameter. The objective function for hyperedge contrast is the average loss across all positive pairs
\vspace{-0.1 in}
\begin{align}
    \mathcal{L}^{\text{ec}} = \frac{1}{2|\mathcal{E}|} \sum^{|\mathcal{E}|}_{n=1} \left( \ell \left(\bm{x}^{\langle 1 \rangle}_{e_{n}}, \bm{x}^{\langle 2 \rangle}_{e_{n}} \right) + \ell \left(\bm{x}^{\langle 2 \rangle}_{e_{n}}, \bm{x}^{\langle 1 \rangle}_{e_{n}} \right) \right).
\end{align}
\textit{Device-hyperedge membership contrast} seeks to distinguish between real and fake device-hyperedge memberships across the two augmented views. For any device $a_i$ and hyperedge $e_n$ that form membership in the original hypergraph, i.e., $a_i \in e_n$, the device embedding from the first view, $\bm{x}_{a_i}^{\langle 1 \rangle}$, is designated as the anchor, and the corresponding hyperedge embedding from the second view, $\bm{x}_{e_n}^{\langle 2 \rangle}$, is considered the positive sample. Negative samples are selected from the embeddings of the other hyperedges that are not associated with device $a_i$. Conversely,  when $\bm{x}_{e_n}^{\langle 2 \rangle}$ is set as the anchor, negative samples are drawn from the embeddings of devices that are not associated with  $e_n$. Therefore, the loss function for a pair of $\bm{x}^{\langle 1 \rangle}_{a_i}$ and $\bm{x}^{\langle 2 \rangle}_{e_n}$ is given by
\begin{align}
    & \ell \left ( \bm{x}^{\langle 1 \rangle}_{a_i},  \bm{x}^{\langle 2 \rangle}_{e_n}\right) = \nonumber\\ 
    & - 
    \underbrace{\log \frac{e^{\mathcal{D}(\bm{x}^{\langle 1 \rangle}_{a_i},  \bm{x}^{\langle 2 \rangle}_{e_n})/ \tau^{\text{mc}}}}{e^{\mathcal{D}(\bm{x}^{\langle 1 \rangle}_{a_i},  \bm{x}^{\langle 2 \rangle}_{e_n})/ \tau^{\text{mc}}} + \sum_{k:a_i \notin e_{k}} e^{\mathcal{D}(\bm{x}^{\langle 1 \rangle}_{a_i},  \bm{x}^{\langle 2 \rangle}_{e_k})/ \tau^{\text{mc}}}} }_{\bm{x}^{\langle 1 \rangle}_{a_i} \text{ is the anchor}}\nonumber \\
     & - \underbrace{\log \frac{e^{\mathcal{D}(\bm{x}^{\langle 1 \rangle}_{a_i},  \bm{x}^{\langle 2 \rangle}_{e_n})/ \tau^{\text{mc}}}}{e^{\mathcal{D}(\bm{x}^{\langle 1 \rangle}_{a_i},  \bm{x}^{\langle 2 \rangle}_{e_n})/ \tau^{\text{mc}}} + \sum_{i:a_i \notin e_{n}} e^{\mathcal{D}(\bm{x}^{\langle 1 \rangle}_{a_i},  \bm{x}^{\langle 2 \rangle}_{e_n})/ \tau^{\text{mc}}}}}_{\bm{x}^{\langle 2 \rangle}_{e_{n}} \text{ is the anchor}},
\end{align}

where $\tau^{\text{mc}}$ is a temperature parameter, and {\footnotesize $\mathcal{D} (\bm{x}^{\langle 1 \rangle}_{a_i},  \bm{x}^{\langle 2 \rangle}_{e_n})$} is a bilinear function used to calculate the probability assigned to this device-hyperedge. To reduce the computational complexity, we randomly select one negative sample for each positive sample in the calculation of $\footnotesize \ell \left ( \bm{x}^{\langle 1 \rangle}_{a_i},  \bm{x}^{\langle 2 \rangle}_{e_n}\right)$. The objective function for the device-hyperedge membership contrast is defined as
\vspace{-0.07 in}
\begin{small}
    \begin{align}
    \mathcal{L}^{\text{mc}} = \frac{1}{2|\mathcal{A}| |\mathcal{E}|} 
  \sum_{i=1}^{|\mathcal{A}|} \sum^{|\mathcal{E}|}_{n=1} \left( \ell \left(\bm{x}^{\langle 1 \rangle}_{a_i}, \bm{x}^{\langle 2 \rangle}_{e_{n}} \right) + \ell \left(\bm{x}^{\langle 2 \rangle}_{a_i}, \bm{x}^{\langle 1 \rangle}_{e_{n}} \right) \right).
\end{align}
\end{small}
Therefore, the total contrastive loss is formulated as
\vspace{-0.06 in}
\begin{align}
    \mathcal{L} =   
    \mathcal{L}^{\text{dc}} + \omega^{\text{ec}} \mathcal{L}^{\text{ec}} + \omega^{\text{mc}} \mathcal{L}^{\text{mc}},
\end{align}
where $\omega^{\text{ec}}$ and $\omega^{\text{mc}}$ are the weights of $\mathcal{L}^{\text{ec}}$ and $\mathcal{L}^{\text{mc}}$, respectively. By minimizing $\mathcal{L}$, the contrastive learning framework outputs the embeddings of devices, $\bm{X}_{\mathcal{A}}$, and the embeddings of hyperedges, $\bm{X}_{\mathcal{E}}$. 

\vspace{-0.06 in}
\subsection{Trust calculation}

After obtaining the embeddings of devices, their trust values can be directly calculated based on these embeddings, as the high-order social relationships between devices are mapped into a space of the same dimensionality through the proposed HSCL approach. The trust value of $a_j$ assessed by $a_i$ is determined using cosine similarity
\vspace{-0.07 in}
\begin{align}
  T_{a_i \to a_j}  = \frac{\bm{x}_{a_i} \cdot \bm{x}_{a_j}}{\parallel \bm{x}_{a_i}\parallel \parallel \bm{x}_{a_j} \parallel}, \bm{x}_{a_i}, \bm{x}_{a_j} \in \bm{X}_{\mathcal{A}}.
\end{align}
Finally, the task initiator $a_i$ evaluates the trust values of all potential collaborators and selects the one with the highest trust value as the trusted collaborator.

\section{Experiments}
\label{simulation}

\subsection{Experimental setup}

\subsubsection{Dataset}
To evaluate the proposed HSCL method, the Sigcomm-2009 dataset is used \cite{b20}, consisting of traces that can be mapped to the promising paradigm of social IoT. These traces include social information related to devices/users, such as friendships, interests, activities, and message logs. The dataset contains 76 nodes and 18,226 interactions over a span of four days. Since the dataset does not include the geographic coordinates of the nodes, we assign each node a coordinate in a two-dimensional space.

\subsubsection{Hyperparameters} 

The argumentation hyperparameters $p^{\mathcal{A}}$, $p^{\mathcal{E}}$, and $p^{\bm{H}}$, which govern the sampling process for masking, are selected within the range of 0.0 to 0.4. Additionally, three temperature hyperparameters $\tau^{\text{dc}}$, $\tau^{\text{ec}}$, and $\tau^{\text{mc}}$, which control the uniformity of the embedding distribution, are chosen from the range of 0.1 to 1.0. The embedding size is set to 512. HGNN is trained using the Adam optimizer with a weight decay of $10^{-5}$. The stiffness parameter $\beta$ in the soft $K$-means is set to 0.4. The proposed model is implemented by Python and PyTorch.

 \begin{figure}[!t]
      \centering
      \subfigure[The trust value distribution calculated by the proposed HSCL method]{\includegraphics[scale=0.54]{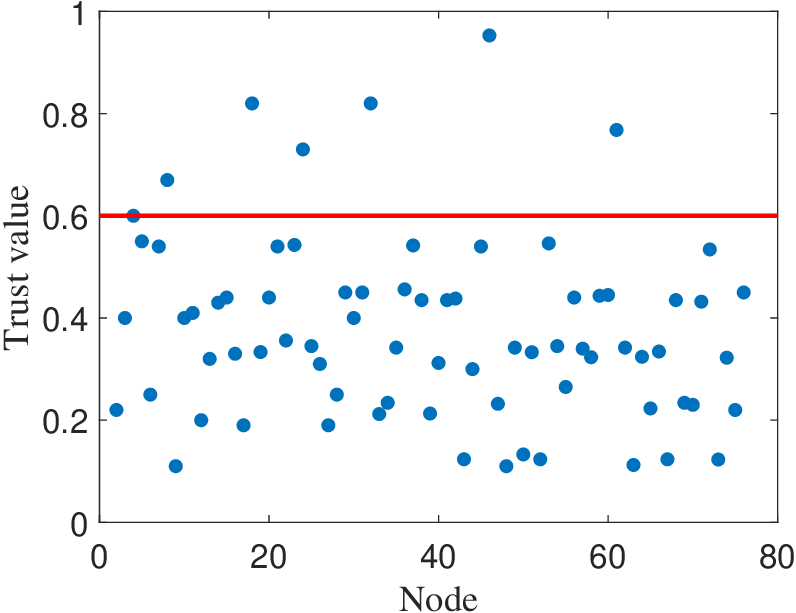}}\hspace{0 in}%
   \subfigure[The trust value distribution calculated by TSTCM]{\includegraphics[scale=0.54]{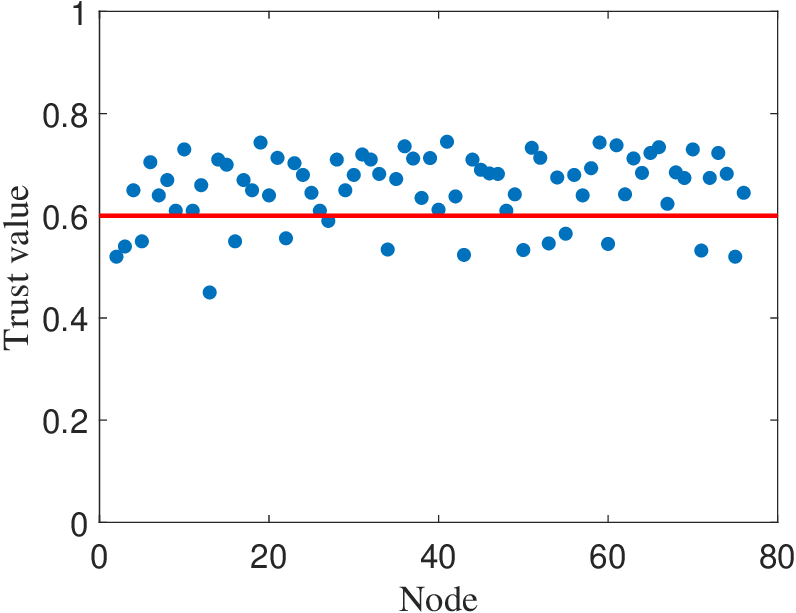}}
     \caption{Comparison of trust values.}
     \label{trust_distribution}
\end{figure}

\subsection{Comparison of trust values}
In this subsection, the time-aware similarity-based trust computational model (TSTCM) \cite{b3} is used as the comparison method. Node 1 serves as the task initiator and evaluates the trust values of the other 75 nodes. The experiment results are shown in Fig.~\ref{trust_distribution}, where the x-axis represents the nodes, the y-axis shows the trust values, and the trust threshold is set at 0.6. From Fig.~\ref{trust_distribution}~(a), it is evident that the proposed HSCL method effectively distinguishes the trustworthiness of the nodes. In other words, nodes trusted by node 1 receive notably higher trust values, whereas untrusted nodes generally receive lower values. In Fig.~\ref{trust_distribution}~(b), however, the TSTCM algorithm does not clearly differentiate the trust levels of nodes, as most trust values fall within a relatively narrow range. This limited differentiation can lead to inaccuracies when selecting trusted nodes for collaboration. Therefore, the proposed HSCL approach outperforms the comparison algorithm in distinguishing between trusted and untrusted nodes.
\subsection{Comparison of selected trustworthy nodes}
\begin{figure}[t!]
	\centering
	\hspace{0 in}\includegraphics[scale=0.335]{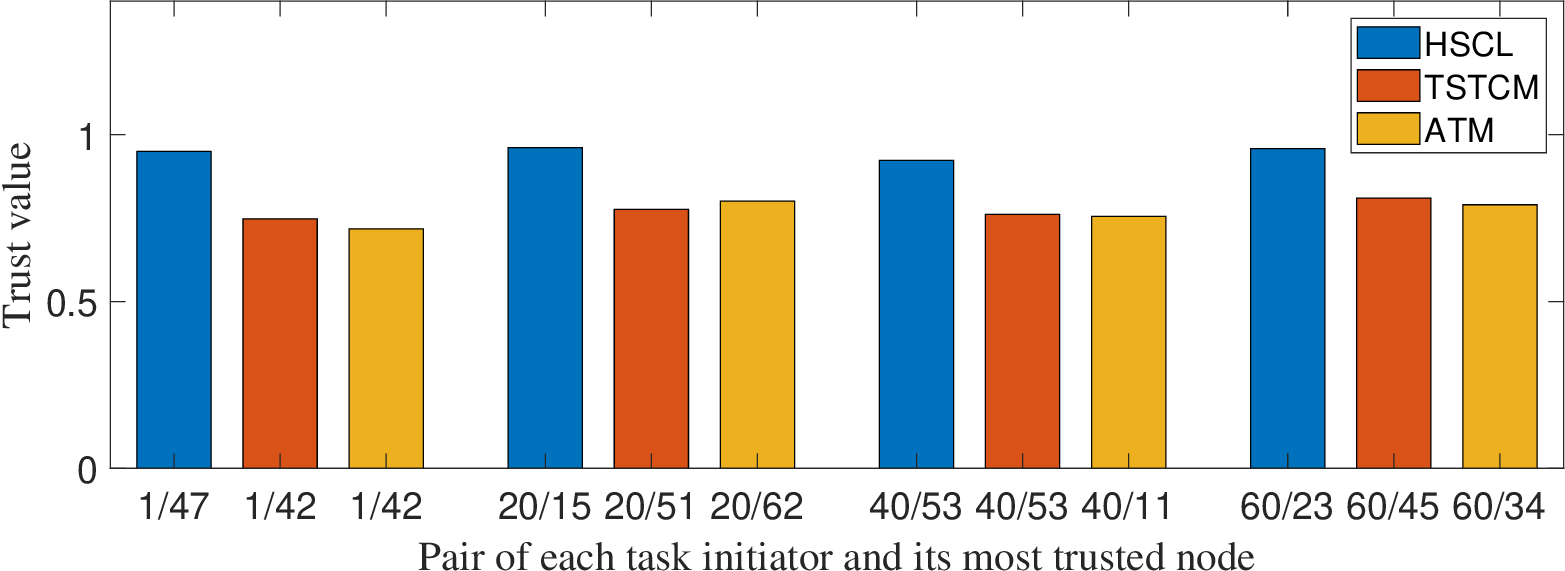}
	\caption{Comparison of selected trustworthy nodes.}
	\label{selected_collaborator}
\end{figure}

In this subsection, the most trusted nodes identified by the proposed HSCL algorithm are compared with those selected by the TSTCM and adaptive trust management (ATM) methods~\cite{b21}. As shown in Fig.~\ref{selected_collaborator}, the x-axis represents pairs of each task initiator and the selected trusted node. For example, `1/47' means that node 1 is the task initiator, and node 47 is the selected node with the highest trust value. We can observe that the most trusted nodes selected by our algorithm have higher trust values compared to those selected by the comparison algorithms. Specifically, when node 40 is the task initiator, HSCL and TSTCM both identify node 53 as the most trusted node. However, the trust value of node 53 calculated by our algorithm is higher than that calculated by TSTCM. Therefore, our method can more accurately calculate the trust values between devices and identify the most trusted node.

\subsection{Comparison of selected trustworthy nodes when changing the number of nodes}
\begin{figure}[t!]
	\centering
	  \includegraphics[scale=0.52]{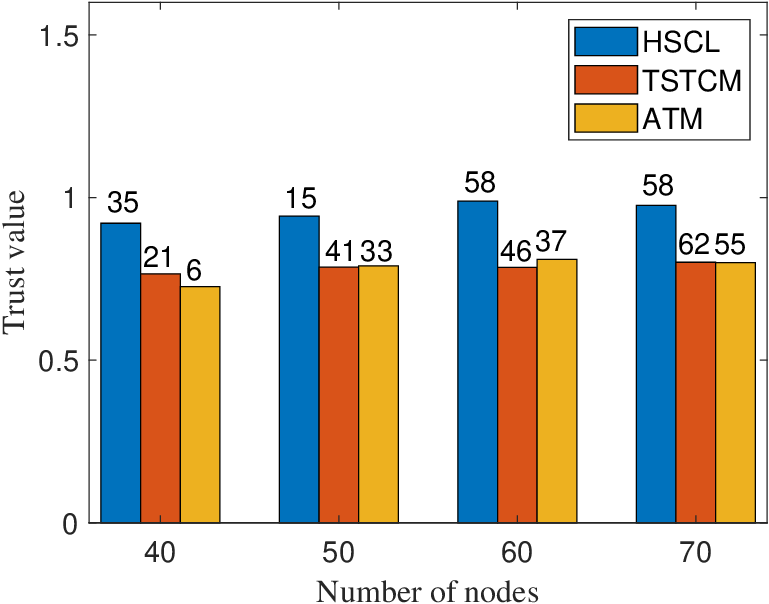}
	\caption{Comparison of selected trustworthy nodes when changing the number of nodes.}
	\label{chaningsize}
\end{figure}

In this subsection, we investigate the impact of the number of nodes on the selection of the most trusted node. With node 10 as the task initiator, the most trusted nodes identified by the three methods are presented in Fig.~\ref{chaningsize}. The x-axis in Fig.~\ref{chaningsize} represents the total number of nodes in the system, while the numbers above the bars indicate the identifiers of the selected most trusted nodes. For example, when the total number of nodes is 40, the most trusted node selected by our proposed HSCL method is node 35, with a trust value of 0.92.  We can observe that as the number of nodes varies, the most trusted nodes selected by the three methods also change. However, the nodes chosen by our proposed HSCL method consistently have the highest trust values. Furthermore, node 58 is selected as the most trusted node by our method when the number is 60 and 70, even though its trust value slightly changes. This indicates that the increase in the number of nodes introduces more social relationships, impacting trust values.


\section{Conclusion}
\label{conclusion}

This paper proposed the novel HSCL method to accurately assess trust between devices in the IoT system with complex social attributes. First, hypergraphs were employed to mine and represent the complex and high-order relationships based on social attributes. To enrich the semantics of the generated social relationship hypergraph, hypergraph augmentation was applied. Furthermore, a parameter-sharing HGNN was used to nonlinearly fuse these high-order social relationships. In addition, a self-supervised contrastive learning method was employed to derive more meaningful device embeddings, which were subsequently used to calculate trust values between devices. Extensive experiments demonstrated that the proposed HSCL method can effectively distinguish between trusted and untrusted nodes, and consistently select the most trusted node compared to baseline algorithms.


\end{document}